Distributional properties of semantic interference in picture naming: Bayesian meta-analyses

Pamela Fuhrmeister*

Audrey Bürki

Department of Linguistics
University of Potsdam
Karl-Liebknecht-Straße 24-25
14476 Potsdam, Germany

*Corresponding author, email: fuhrmeister@uni-potsdam.de

Abstract

Studies of word production often make use of picture-naming tasks, including the picture-word-interference task. In this task, participants name pictures with superimposed distractor words. They typically need more time to name pictures when the distractor word is semantically related to the picture than when it is unrelated (the semantic interference effect). The present study examines the distributional properties of this effect in a series of Bayesian meta-analyses. Meta-analytic estimates of the semantic interference effect first show that the effect is present throughout the reaction time distribution and that it increases throughout the distribution. Second, we find a correlation between a participant's mean semantic interference effect and the change in the effect in the tail of the reaction time distribution, which has been argued to reflect the involvement of selective inhibition in the naming task. Finally, we show with simulated data that this correlation emerges even when no inhibition is used to generate the data, which suggests that inhibition is not needed to explain this relationship.

Distributional properties of semantic interference in picture naming: Bayesian meta-analyses

The cognitive processes underlying word production are often assessed using picture-naming tasks. In one such task, the picture-word interference task, participants are asked to name a picture in the presence of a superimposed distractor word (see Figure 1). Participants typically take longer to name a picture when the distractor word is semantically related to the picture than when the distractor is unrelated (semantic interference effect, e.g., Lupker, 1979; Bürki et al., 2020). Findings from this paradigm have been used to inform a variety of issues, including the relationship between linguistic processes and other cognitive functions (e.g., Shao et al., 2013). Specifically, the distributional properties of the semantic interference effect have been argued to inform the mechanisms underlying the effect and involvement of cognitive abilities, such as attention or inhibition. The present study presents a series of Bayesian meta-analyses targeting different aspects of the distributional properties of the semantic interference effect. Our first aim is to provide estimates of the magnitude of the effect at different points in the distribution. Our second aim is to examine the relationship between the magnitude of the semantic interference effect and the change in effect size in slow response times, a relationship assumed to reflect the involvement of selective inhibition.

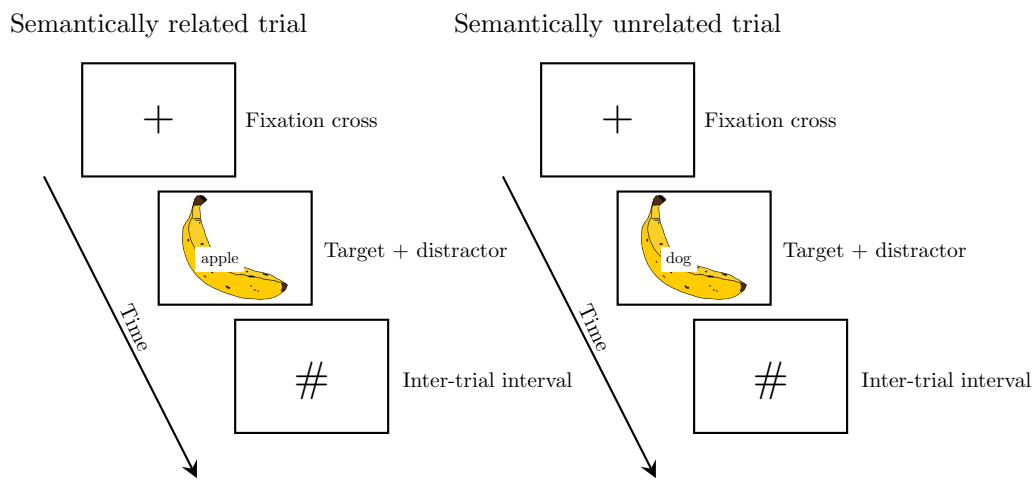

*Figure 1. Examples of trials in a picture-word-interference task. Participants typically need more time to name the picture in the presence of semantically related distractor words (left) than semantically unrelated distractor words (right).*

Distributional analyses examine how an experimental effect evolves with the response time distribution (e.g., Balota & Yap, 2011). One way this is examined is by using delta or Vincintile plots (Balota & Yap, 2011). To produce these plots, reaction times for each trial are first rank ordered per participant and condition and divided into quantiles (percentile bins). Mean reaction times are computed for each condition in each quantile and the difference between the two conditions (delta) is plotted for each quantile, which shows how the effect changes over the response time distribution (see Figure 2). Studies that have examined distributional properties of the semantic interference effect in this way all agree that semantic interference increases with increasing reaction times (e.g., Roelofs & Piai, 2017). Some studies find the effect across the entire response time distribution; in others, the effect is restricted to the slowest quantiles of the distribution (Scaltritti et al., 2015). This distributional pattern has been related to attention. It has been argued, for example, that lapses of attention generate both longer naming times and greater interference from the distractor word (De Jong, 1999; see also discussion in Roelofs, 2008). According to Van Maanen and Van Rijn (2008), an increase in the effect in the tail of the distribution could come from a restricted number of trials where the distractor was wrongly selected as a response, an error that is then corrected so that the correct response can be selected. In the present study, we take advantage of multiple datasets to determine whether the semantic interference effect is present over the entire distribution of response times or restricted to slow responses.

In classical conflict tasks such as the Stroop or Flanker task, changes in effect size toward the end of the distribution have been related to selective inhibition abilities (Proctor et al., 2011; Ridderinkhof et al., 2004, 2005; van den Wildenberg et al., 2010). In such tasks, participants respond to a stimulus on each trial, and some trials provide congruent information, and some provide incongruent information that needs to be inhibited. According to the activation suppression hypothesis, (Ridderinkhof et al., 2004, 2005) inhibition takes time to build up, making it more effective in trials with slower reaction times than faster trials. Interference or congruency effects (i.e., the difference in reaction times between incongruent and congruent trials) are expected to increase with reaction times (Figure 2A); however, when inhibition is applied, the difference between congruent and incongruent trials tends to decrease with increasing reaction times and can even become negative (i.e., a facilitation effect, Figure 2B). The slope (or change in effect size) for the last delta segment derived following the delta or Vincintile procedure described above (e.g., the line between the slowest two quantiles in Figure 2) has been taken as a measure of selective inhibition and has been used with different tasks and populations (e.g., Ridderinkhof et al., 2005). Interestingly for our purposes, studies have applied this procedure to measure individual or group differences in inhibition deployed in the picture-word-interference task when pictures are presented with semantically related or unrelated distractor words.

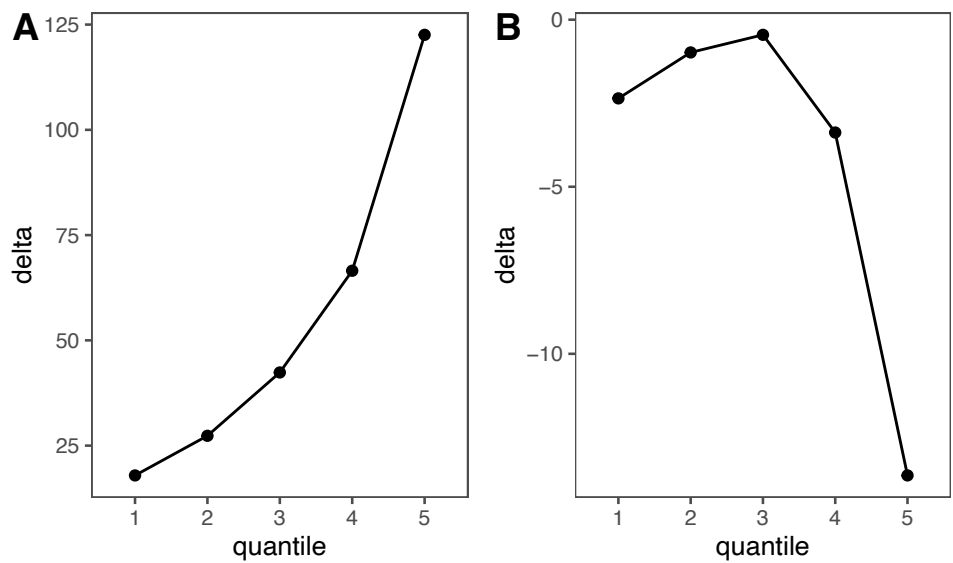

*Figure 2. Example distributional patterns of the semantic interference effect calculated over five quantiles. A. A positive slope between the last two quantiles, indicating the effect increased over the slowest trials, and B. a negative slope between the last two quantiles, indicating the interference effect turned into a facilitation effect over the slowest trials. Note the difference in scales on the y-axis.*

For example, Shao et al. (2013) found that a participant's slope of the slowest delta segment was correlated with the magnitude of the semantic interference effect: Participants who had a less positive slope for the slowest delta segment showed a smaller semantic interference effect overall. This finding was replicated and extended to another task (the semantic blocking task) in Shao et al. (2015), and a similar pattern has been found in bilingual participants (Roelofs et al., 2011). These findings were taken to reflect inter-individual variability in the ability to apply selective inhibition. According to this view, participants with larger effects are less able to apply selective inhibition than participants with smaller effects.

In at least one analysis, Shao et al. (2013) observed a similar relationship by item. For this analysis, the mean by-item semantic interference effect was correlated with the change in effect size in the last delta segment, which Shao et al. (2013) took to further confirm the inhibition explanation. In a within-participant design, all participants name all items in both conditions. This finding is therefore interesting because it suggests that differences in inhibition are also visible, for a given participant, across trials.

According to the activation suppression hypothesis, inhibition requires time to build up. Shao et al. (2015) reasoned that under this hypothesis, no correlation should be found with the fastest segment. Shao et al. (2015) found that the slope of the *fastest* delta plot segment was correlated with the magnitude of the semantic interference effect in one of two picture-word-interference experiments (but not in two experiments using a semantic blocking task). The fact that they saw a relationship with the slope of the *slowest* delta segment but not with the *fastest* (with the exception of the one experiment) was taken to lend support to claims made by the activation suppression hypothesis by Ridderinkhof et al. (2004)—that inhibition takes time to build up and is mostly reflected in the slope of the slowest delta segment. We note that if the effect increases with response times, more power is likely necessary to detect a correlation in earlier segments, where the effect is smaller. More evidence is therefore needed to determine whether only the slope of the slowest delta segment reflects selective inhibition. With a meta-analysis, the chances of detecting even small effects are increased. In the present study, we present meta-analyses of the correlation between the magnitude of the semantic interference effect and the increase in effect size in the first and last delta segments.

## Meta-analyses

We report several meta-analyses (see Table 1) examining the distributional properties of the semantic interference effect. In all studies included in the meta-analysis, participants performed a picture-word-interference task: They named pictures while ignoring distractors either semantically related or unrelated to the target. Meta-analyses provide information on the reliability of a pattern across datasets and provide estimates of effect sizes and their uncertainty. On top of informing theoretical issues, estimates of effect sizes and of uncertainty are useful to calculate power for subsequent studies.

Our first set of meta-analyses tests whether the semantic interference effect is present across the whole reaction time distribution or only in the slower portion, and it provides estimates of the effect for different parts of the distribution. Our second set of meta-analyses provides estimates of the correlation between the mean effect size of the semantic interference effect and the effect size in the slowest and fastest segments, both by participant and by item.

To anticipate, our analyses provide evidence that the semantic interference effect is present over the entire distribution and increases with response times. They also confirm the correlation between effect size and slowest as well as fastest segments, both by participant and by item. These results lead us to consider an alternative account of this relationship, which we examine using simulations.

| Effect of interest | Quintiles calculated by participants or items |
| --- | --- |
| 1. Magnitude of the semantic interference effect in first quintile<br>2. Magnitude of the semantic interference effect in second quintile<br>3. Magnitude of the semantic interference effect in third quintile<br>4. Magnitude of the semantic interference effect in fourth quintile<br>5. Magnitude of the semantic interference effect in fifth quintile | Participants |
| 6. Correlation between the magnitude of the semantic interference effect and the slowest delta plot segment | Participants |
| 7. Correlation between the magnitude of the semantic interference effect and the fastest delta plot segment | Participants |
| 8. Correlation between the magnitude of the semantic interference effect and the slowest delta plot segment | Items |
| 9. Correlation between the magnitude of the semantic interference effect and the fastest delta plot segment | Items |

*Table 1. Summary of meta-analyses: effects of interest and whether quintiles were calculated from participant or item data.*

**Methods**

*Data set*

We worked with a subset of data collected for a previous meta-analysis of the semantic interference effect (Bürki et al., 2020). We selected all the studies for which we had the raw data (a response time for each trial) and that had a stimulus onset asynchrony (SOA) between -160 and 160 ms. Several studies have reported semantic interference effects in this SOA range (e.g., Damian & Martin, 1999; Glaser & Düngelhoff, 1984; Starreveled & La Heij, 1996), a pattern supported by a recent meta-analysis (Bürki et al., 2020). We included two additional data sets that were recently collected from our lab.

Participants in all studies were adult native speakers of the language being tested, and they did not have language disorders. Languages tested in the various studies included German, English, French, Italian, Dutch, Spanish, and Mandarin. Only trials with distractor items that were semantically related or unrelated to the target picture were considered. Multiple experiments within a paper were treated as independent datasets. Experiments where the same items were tested at different SOAs or with and without familiarization were split to generate one dataset for each level of these variables. This resulted in a total of 54 datasets from 22 different experiments. More details on these studies can be found in Appendix B.

*Extraction of estimates*

Only correct responses were included in the analyses. Reaction time data were first separated by participant and condition (semantically related or unrelated trials). Reaction times were then sorted and divided into quantiles and the mean difference between conditions for each quantile was computed. We used five quintiles (i.e., 20% bins) as in Shao et al. (2013, 2015).

A total of nine participants from all data sets were eliminated because they did not have enough data points to calculate quantiles.

*Estimates of semantic interference in each quintile.* For each study, we fit a linear mixed effects model using the lme4 package (Bates et al., 2015) in R (R Core Team, 2020). Each model predicted naming latencies (the dependent variable) and included fixed effects of quintile (1-5), and condition (deviation coded, semantically related = .5, semantically unrelated = -.5), which was nested within quintile. Nested fixed effects allow us to test for simple effects (Schad et al., 2020), and in this case, we were interested in testing the difference in reaction times between semantically related and unrelated conditions at each level of the factor quintile (i.e., in each quintile separately). Random effects included by-participant random intercepts and slopes for quintile and by-participant random intercepts and slopes for condition, which were nested within quintile. Item random effects included by-item random intercepts and slopes for quintile and random intercepts and slopes for condition, which were nested within quintile. Correlations between random effects were set to zero.

*Estimates of correlations.* We first calculated the mean semantic interference effect as well as the slope for the slowest delta segment (i.e., the slope between quintiles four and five) and fastest delta segments (the slope between quintiles one and two) for each participant. We followed the procedure in, for example, De Jong (1994), Ridderinkhof et al. (2004), Roelofs et al., (2011), and Shao et al. (2013) to calculate the slope as follows:

slope(quintile 4, quintile 5) = (delta(quintile 5) - delta(quintile 4))/(mean(quintile 5) - mean(quintile 4))

For each study, we computed the correlation between the slope of the *slowest* delta segment and the mean semantic interference effect and also for the slope of the *fastest* delta segment and the mean semantic interference effect. We then used the Fisher z-transformation to transform correlation coefficients ($r$ values) to $z$ values (Fisher, 1915) using the FisherZ function from the DescTools package (Signorell et al., 2020) in R. The z-transformed scores and their estimated standard error were entered into the meta-analyses described below. The whole process was repeated for by-item analyses.

### *Meta-analyses.*

Meta analyses estimate the size and uncertainty of an effect in question from the effect sizes and standard errors of individual studies. Both fixed-effects and random-effects meta-analyses can be performed, but they make different assumptions. Fixed-effects meta-analyses assume that all studies have the same true effect $\theta$ (e.g., Chen & Peace, 2013), but random-effects meta-analyses assume that the different studies have different true effects $\theta_i$ (e.g., Sutton & Abrams, 2001). Each of the studies included in our data set were performed in different languages and in different labs; therefore, we assume a different underlying effect for each study and thus performed a random-effects meta-analysis.

For meta-analyses testing the magnitude of the semantic interference effect in each quintile, we made the following assumptions: Each study $i$ has a true effect of $\theta_i$ that is normally distributed with a mean of $\theta$ and variance of $\tau^2 = 100^2$. The observed effect of the predictor $y_i$ in each study is assumed to stem from a normal distribution with mean $\theta_i$ and variance $\sigma_i^2$, the true standard error of the study. Details of the model specifications can be found in Equations (1).

$$y_i | \theta_i, \sigma_i^2 \sim N(\theta_i, \sigma_i^2) \; i = 1, \dots, n,$$
$$\theta_i | \theta, \tau^2 \sim N(\theta, \tau^2),$$
$$\theta \sim N(0, 100^2), \quad (1)$$
$$\tau \sim N(0, 100), \tau > 0$$

$y_i$ represents the observed effect of the predictor in each study $i$; $\theta$ is the true effect of the predictor estimated by the model; $\sigma_i^2$ represents the variance for study $i$, estimated from the standard error of the effect of the predictor for this study; and $\tau^2$ represents the between-study variance.

For the meta-analyses testing the magnitude of the semantic interference effect in each quintile, we chose weakly informative priors from a normal distribution with a mean of 0 and a standard deviation of 100. For the standard deviation, we chose weakly informative priors from a truncated normal distribution with a mean of 0 and standard deviation of 100.

For the meta-analyses of correlations (note that the meta-analysis is performed on the Fisher z-transformed correlations: assumptions and prior pertain to the z-transformed score), we assumed the following: Each study $i$ has a true z-transformed correlation of $\zeta_i$ that is normally distributed with a mean of $\zeta$ and variance of $\tau^2 = 10^2$. The observed z-transformed correlation $z_i$ in each study is assumed to stem from a normal distribution with mean $\zeta_i$ and variance $\phi_i^2$, the true standard error of the study. Details of the model specifications can be found in Equations (2).

$$z_i | \zeta_i, \phi_i^2 \sim N\left(\zeta_i, \phi_i^2\right) \, i = 1, \dots, n,$$
$$\zeta_i | \zeta, \tau^2 \sim N(\zeta, \tau^2),$$
$$\zeta \sim N(0, 10^2), \tag{2}$$
$$\tau \sim N(0, 10), \tau > 0$$

$z_i$ represents the observed z-transformed correlation in each study $i$; $\zeta$ is the true z-transformed correlation estimated by the model; $\phi_i^2$ represents the standard error for this study; and $\tau^2$ represents the between-study variance.

For the intercept and standard deviation for meta-analyses of correlations, we chose weakly informative priors from a normal distribution with a mean of 0 and standard deviation of 10. For the standard deviation, we chose weakly informative priors from a truncated normal distribution with a mean of 0 and standard deviation of 10.

We additionally did sensitivity analyses for each meta-analysis. Effect sizes did not change with different priors for any of the meta-analyses, and details on sensitivity analyses can be found in Appendix A. We also did Bayes factor tests to test whether we have relative evidence for the effect (the alternative hypothesis) over the null hypothesis. Bayes factors of one indicate no evidence, and Bayes factors greater than 10 or less than 1/10 are typically considered to reflect "substantial evidence" for one model over the other (e.g., Wetzels & Wagenmakers, 2012). Meta-analyses were performed in R (R Core Team, 2020) with the brms package (Bürkner, 2018). Data and analysis code can be found at https://osf.io/v2fx5/.

**Results**

Results of the meta-analyses are summarized in Tables 2-4. Meta-analytic estimates, tau (standard deviation), their 95% credible intervals (CrI), and Bayes factors in favor of the alternative hypothesis ($BF_{10}$) are reported.

| Meta-analysis: by-participant analyses | Estimate | 95% CrI | tau | 95% CrI | $BF_{10}$ |
|---|---|---|---|---|---|
| Semantic interference effect in quintile 1 | 6 ms | [3, 9] | 2 ms | [0, 5] | 585 |
| Semantic interference effect in quintile 2 | 11 ms | [8, 14] | 6 ms | [3, 9] | 22037195 |
| Semantic interference effect in quintile 3 | 17 ms | [13, 21] | 10 ms | [7, 14] | 2005103657 |
| Semantic interference effect in quintile 4 | 27 ms | [22, 32] | 15 ms | [11, 20] | 370895977126 |
| Semantic interference effect in quintile 5 | 49 ms | [39, 58] | 22 ms | [12, 32] | 230876883964 |

*Table 2. Results of meta-analyses testing the magnitude of the semantic interference effect per quintile (see Figure 3).*

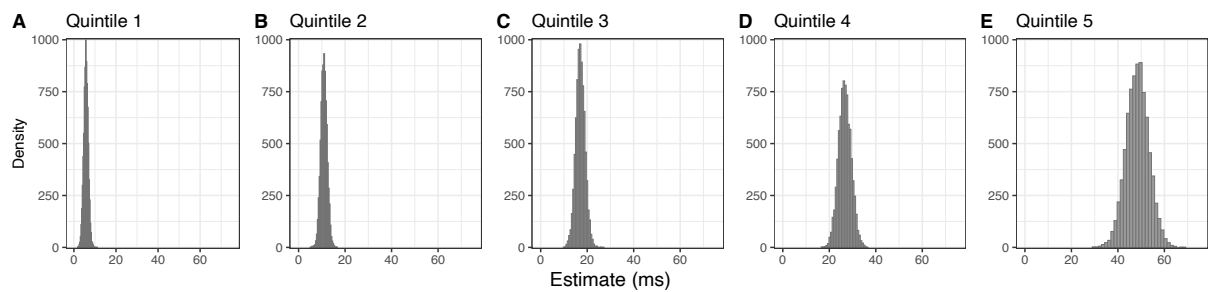

*Figure 3.* Posterior distributions of meta-analytic estimates of the semantic interference effect in each quintile. Quintiles were calculated by participant.

| Meta-analysis: by-participant analyses | Estimate | CrI | tau | CrI | BF$_{10}$ |
|---|---|---|---|---|---|
| Correlation between *slowest* delta plot segment slope and semantic interference effect | .52 | [.44, .61] | .23 | [.16, .31] | 114098240351560 |
| Correlation between *fastest* delta plot segment slope and semantic interference effect | .18 | [.11, .24] | .13 | [.04, .22] | 131 |

*Table 3.* Results of meta-analyses testing the relationships between the semantic interference effect and the slowest and fastest delta plot segments when quintiles are calculated by participant. Estimates for correlational meta-analyses are Fisher z-transformed units; however, r and Fisher z-transformed values are very similar for r values between -.5 and .5 (see Figure 4).

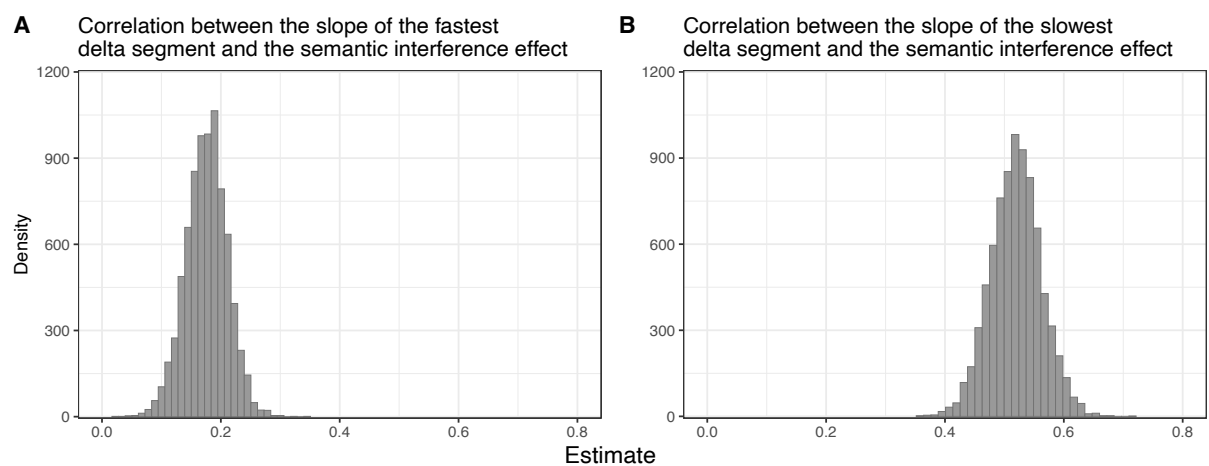

*Figure 4.* Posterior distributions of meta-analytic estimates (Fisher z-transformed correlations) of the relationship between A. the semantic interference effect and the slope of the fastest delta segment and B. the semantic interference effect and the slope of the slowest delta segment. Quintiles were calculated by participant.

| Meta-analysis: by-item analyses | Estimate | CrI | tau | CrI | BF$_{10}$ |
|---|---|---|---|---|---|
| Correlation between *slowest* delta plot segment slope and semantic interference effect | .49 | [.43, .54] | .07 | [.00, .15] | 5410719887112496128 |
| Correlation between *fastest* delta plot segment slope and semantic interference effect | .31 | [.26, .36] | .04 | [.00, .12] | 2944056121802 |

Table 4. Results of meta-analyses testing the relationships between the semantic interference effect and the slowest and fastest delta plot segments when quintiles are calculated by item. Estimates for correlational meta-analyses are Fisher z-transformed units; however, r and Fisher z-transformed values are very similar for r values between -.5 and .5 (see Figure 5).

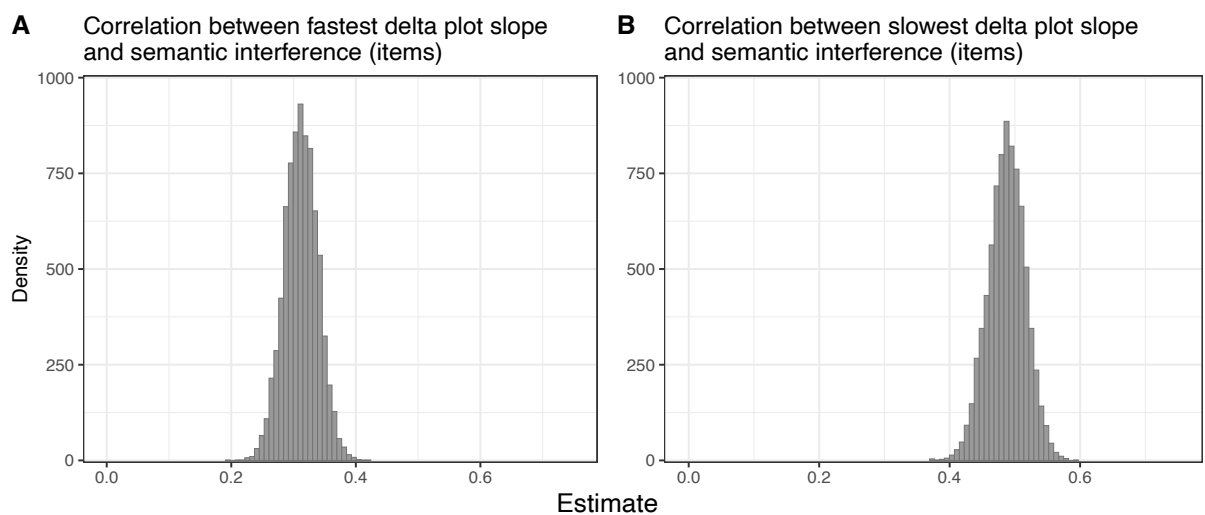

Figure 5. Posterior distributions of meta-analytic estimates (Fisher z-transformed correlations) of the relationship between A. the semantic interference effect and the slope of the fastest delta segment and B. the semantic interference effect and the slope of the slowest delta segment. Quintiles were calculated by items rather than participants.

### Discussion

Our analyses suggest that the semantic interference effect increases with response times[1].

Several mechanisms have been put forward to explain this increase, for example, fluctuations

of attention (Roelofs & Piai, 2017; Scaltritti et al., 2015), selection of wrong responses (Van

---

[1] We additionally confirmed this increase with a Bayes factor test. To do this, we computed the meta-analytic estimate of the interaction between quantile (1-5) and the semantic interference effect and compared this to a null model (BF$_{10}$ = 288485889716).

Maanen & Van Rijn, 2008), or differences in temporal alignment between the processing of the distractor and the encoding of the target word (Bürki & Madec, submitted). We further found that the effect is indeed present even at the fastest response times (the first quintile of the distribution), which suggests that the effect is not solely due to trials where participants selected the wrong response. The meta-analytic estimate of the semantic interference effect in the first quintile is only 6ms; therefore, it is unsurprising that some studies have not found this very small effect, as it will likely require a much larger sample size to detect it.

Interestingly, the increase in the effect size increases with each quintile, with increases of 5ms, 6ms, 10ms and 22ms, respectively. At first sight, this pattern does not seem to fit well with the hypothesis that at least a subset of participants applies more inhibition in the slower quintiles. However, we cannot rule out the possibility that without inhibition, the increase in the magnitude of the effect in the last two quintiles would have been even greater.

The next meta-analysis showed that participants' slopes of the slowest delta segment are positively correlated with the magnitude of their semantic interference effects. This is in line with the hypothesis that participants who apply less inhibition during the task as indexed by a steeper, more positive slope, show larger semantic interference effects. Our analyses confirm the same correlation when we calculated quintiles by item instead of by participant. In a within-item design, each participant names both related and unrelated trials for the same pictures. As a consequence, this correlation cannot *only* be due to an individual-specific ability to apply inhibition but suggests that the slope of the last delta segment also captures *intra*-individual variability in selective inhibition.

We additionally found a correlation between the semantic interference effect and the *fastest* delta segment. According to the activation suppression hypothesis, inhibition takes time to build up, which is why the slope of the *slowest* delta segment is typically used to index inhibition (e.g., Ridderinkhof et al., 2004). Although we found a positive relationship between the slope of the *fastest* delta segment and the semantic interference effect, it was smaller than the size of the correlation with the slowest delta segment. It is possible that with higher-powered studies or with a meta-analysis, we will see that delta slopes start leveling off much earlier. We note that this finding is not necessarily inconsistent with the activation suppression hypothesis because the effect size of the relationship with the slowest delta segment slope was larger than with the fastest segment. It could be argued that inhibition takes time to build up but this does not mean that it is completely absent at shorter response times. Shorter response times likely correspond to words that can be named much more quickly and in the time course of word production, on some trials, inhibition had already had time to build up (response times ranged from 400 to 2000ms).

In the following paragraphs, we consider an alternative explanation to these correlations. Notably, the data in the last (or first) two quantiles are used in both the computation of the mean effect size for the participant (or item) and the computation of the slope of each delta segment. In other words, a correlation may be expected even in the absence of inhibition simply because some of the same data are used to compute the two measures that are then correlated with one another. Moreover, given the increase in effect size throughout the response time distribution, the correlation can be expected to be higher for delta segments where the effect is larger. As the semantic interference effect becomes greater between the last two quintiles, participants who show less of an effect overall could be expected to show less of an increase in the last two quintiles, irrespective of whether they deploy inhibition or not. The question therefore arises of whether these correlations reflect something in addition to this, or whether they are only a by-product of circularity in the procedure[2].

## Simulations

The mechanistic explanation assumes that the increase in effect size with response times combined with the fact that the data in the last two quantiles are used twice in the correlation suffices to generate the correlations we observe. The inhibition account assumes that there is an additional mechanism at play (the application of inhibition on some trials/by some participants). In the first account, negative slopes are due to natural variation, whereas in the inhibition account, they are due to an additional mechanism of inhibition. The goal here is to simulate data under the assumption that no inhibition was applied to test whether the patterns we see in the meta-analysis are due to something more than what we would expect from distributional properties of the semantic interference effect alone.

## Methods

We chose to simulate data generated from an ex-Gaussian distribution. An ex-Gaussian distribution is a convolution of a normal and an exponential distribution and provides a good fit to most reaction time data (Balota & Yap, 2011). Ex-gaussian distributions consist of three parameters: $\mu$, $\sigma$, and $\tau$. $\mu$ and $\sigma$ represent the mean and standard deviation of the normal portion of the distribution, respectively, and $\tau$ represents the mean of the exponential portion of the distribution. As we demonstrated in the first set of meta-analyses, the size of the semantic interference effect increases throughout the response time distribution, a pattern that is produced when the variance of the slower condition (in this case, semantically related trials) is set to be larger than the variance of the faster condition (unrelated trials, Bürki & Madec, submitted; Pratte et al., 2010; Zhang & Kornblum, 1997). We therefore simulated data with greater variance in the slower condition. The code to reproduce the simulations can be found at https://osf.io/v2fx5/.

---

[2] A reviewer mentioned that an argument against a purely mechanistic account is the fact that some participants show a negative slope in the last delta segment or negative deltas in the last quintile, such as in Shao et al. (2013, e.g., Figure 3; 2015). In the context of the selective inhibition account, negative slopes signal a greater amount of inhibition. We note, however, that negative slopes are also expected under the mechanistic account. Most experimental effects in the language production literature, and this is also true for the semantic interference effect, show variability across participants (as can be seen in the variance associated with by-participant random slopes). As a result, a subset of participants shows no effect; another subset shows effects in the other direction. Here, many participants with negative slopes have a negative effect or close to no effect. Again, if the effect is most visible in slower quantiles, it is expected that participants with no or negative effects overall will show flat or negative slopes. Notably, however, if a subset of participants with negative slopes are expected under a mechanistic account, the present data do not allow us to determine whether these are also partly driven by the application of inhibition. In order to shed light on this issue, we conducted simulations.

**Data generation**

We simulated data to mimic a well-powered picture-word-interference experiment with 100 participants, 50 items, and two within participant and within item conditions (semantically related and unrelated distractors, see Bürki et al., 2020 for power analyses for this experimental design). To obtain realistic ex-Gaussian parameter estimates, we pooled data from the studies used in the meta-analyses and estimated the ex-Gaussian parameters for each condition (related and unrelated) separately using the mexgauss() function in the retimes package (Massidda, 2013). We obtained the following estimates for the related condition in milliseconds: $\mu = 578$, $\sigma = 68$, and $\tau = 219$; and for the unrelated condition: $\mu = 570$, $\sigma = 53$, and $\tau = 202$. Ex-Gaussian distributions were simulated for each condition separately with these parameters, and by-participant random intercept adjustments, by-item random intercept adjustments, and residual error were added to the $\mu$ parameter for each simulated trial. Random by-participant intercept adjustments were generated from a normal distribution with mean 0 and standard deviation of 100, by-item random intercepts were generated from a normal distribution with a mean of 0 and standard deviation of 70 (estimated from a previous data set, Fuhrmeister et al., submitted), and residual error was generated from a normal distribution with mean 0 and standard deviation of 100.

## Analysis approach

For each participant and item of the simulated data, we calculated the slope of the first and last delta segments as described in the meta-analyses, and we correlated these values with the mean interference effect for participants and items.

## Results and Discussion

Results of the simulated correlations are summarized in Table 5 and depicted in Figure 6.

| Correlation | Participants/items | *r* | *p*-value |
|---|---|---|---|
| *Slowest* delta segment slope and semantic interference effect | Participants | .46 | <.001 |
| *Fastest* delta segment slope and semantic interference effect | Participants | .29 | .003 |
| S*lowest* delta segment slope and semantic interference effect | Items | .49 | <.001 |
| F*astest* delta segment slope and semantic interference effect | Items | .32 | .02 |

*Table 5. Correlations for each analysis of the simulated data.*

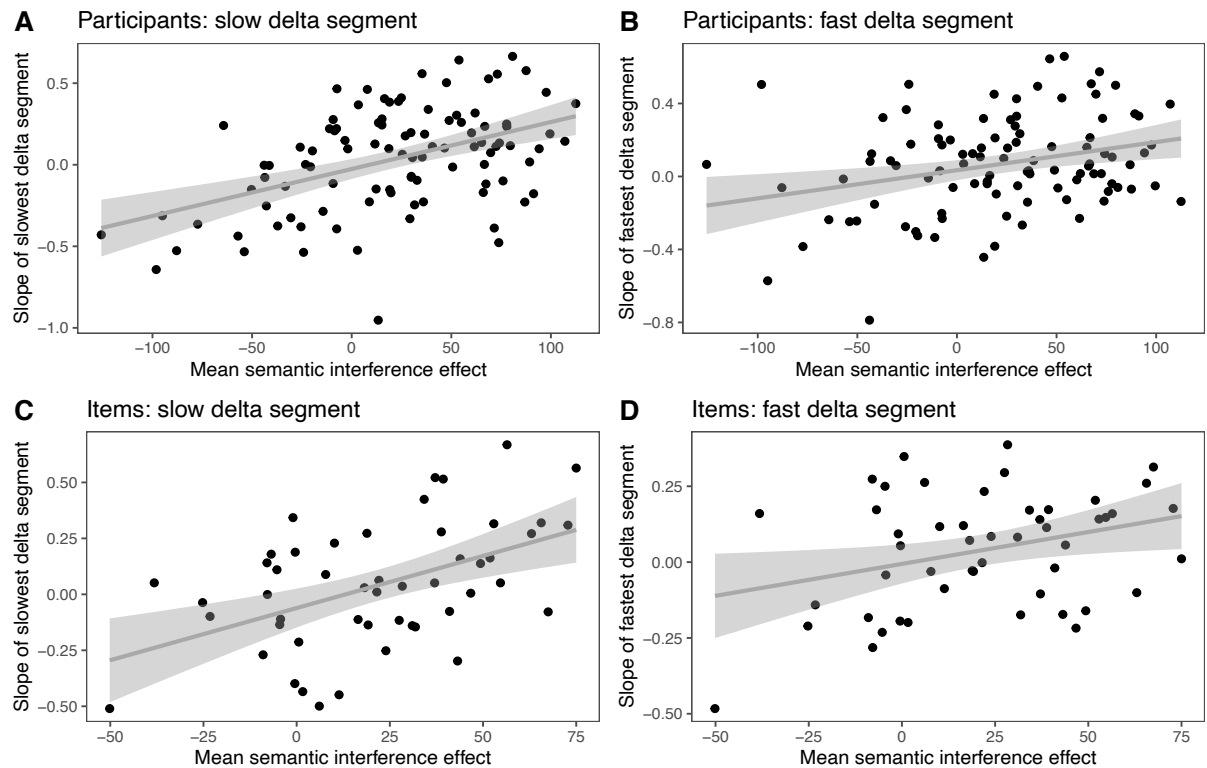

*Figure 6. Correlations from simulated data between the mean semantic interference effect and A. the slope of the slowest delta segment (participants), B. slope of the fastest delta segment (participants), C. slope of the slowest delta segment (items), and D. slope of the fastest delta segment (items).*

The stimulated data suggest that even when we assume no mechanism of inhibition, a similar pattern of correlations emerges between the mean semantic interference effect and the slope of the fastest and slowest delta segments. As can be seen in Figure 6, we additionally see some negative slopes, even when we assume no mechanism of inhibition.

## Discussion

The correlation between the slowest delta segment and the mean effect has been argued to reflect individual differences in the ability to deploy inhibition. However, we showed with simulated data that this relationship is found even when no inhibition is assumed (i.e., when participants did not generate the data), suggesting an inhibition account is not needed to explain this relationship. To be clear, we have not demonstrated that inhibition is not involved in the picture-word-interference task, nor that the slope of the last delta segment does not reflect inhibition. We have simply demonstrated that the correlation between the mean interference effect and the delta segments can also be produced from simulated ex-Gaussian distributions, in which the sigma and tau parameters are larger for the slower condition (see also Bürki & Madec, submitted; Pratte et al., 2010; Zhang & Kornblum, 1997).

It is a reasonable assumption that inhibition is involved in a picture-word-interference task. To conclude that the slope of the last delta segment reflects inhibition in picture-naming tasks, we would need some evidence that is not taken from a correlation with some of the same response times. Shao et al. (2014) provide such evidence in a study in which they asked participants to name pictures with low and high name agreement (the number of different names people assign to an object). The authors reasoned that pictures with low name agreement, for which many potential candidate names would be activated, should require more inhibition than pictures with high name agreement, which have fewer or maybe only one possible name. They calculated delta slopes for this task and found that for action words but not object words, these correlated with the difference in amplitude of the N2 between high and low agreement words, and the N2 is thought to index inhibition. This was taken as evidence that delta slopes in picture-naming tasks reflect inhibition. This correlation suggests that the last delta segment may indeed reflect inhibition; note, however, that this reasoning implies that the difference in the N2 between conditions reflects a measure that is independent of the reaction time difference between the two conditions. In any case, our simulations suggest that the correlation between the slope of the last delta segment and the mean semantic interference effect is there regardless of inhibition.

Acknowledgements

The authors would like to thank all the authors who shared their datasets. This research was funded by the Deutsche Forschungsgemeinschaft (DFG, German Research Foundation) – project number 317633480 – SFB 1287, Project B05 (Bürki).

Open practices statement

The data and materials for all experiments are available at https://osf.io/v2fx5/ and the analyses were not preregistered.

References

Aristei, S., & Abdel Rahman, R. (2013). Semantic interference in language production is due to graded similarity, not response relevance. *Acta Psychologica*, *144*(3), 571–582. https://doi.org/10.1016/j.actpsy.2013.09.006

Aristei, S., Melinger, A., & Abdel Rahman, R. (2011). Electrophysiological chronometry of semantic context effects in language production. *Journal of Cognitive Neuroscience*, *23*(7), 1567–1586. https://doi.org/10.1162/jocn.2010.21474

Balota, D. A., & Yap, M. J. (2011). Moving beyond the mean in studies of mental chronometry: The power of response time distributional analyses. *Current Directions in Psychological Science*, *20*(3), 160-166. https://doi.org/10.1177/0963721411408885

Bates, D. Maechler, M., Bolker, B., & Walker, S. (2015). Fitting Linear Mixed-Effects Models Using lme4. Journal of Statistical Software, 67(1), 1-48. doi:10.18637/jss.v067.i01.

Bürki, A., Elbuy, S., Madec, S., & Vasishth, S. (2020). What did we learn from forty years of research on semantic interference? A Bayesian meta-analysis. *Journal of Memory and Language*, *114*, 104125.

Bürki, A. & Madec, S. (submitted). Picture-word interference in language production studies: Exploring the roles of attention and processing times.

Bürkner, P. C. (2018). Advanced Bayesian Multilevel Modeling with the R Package brms. The R Journal, 10(1), 395–411.

Chen, D. G. D., & Peace, K. E. (2013). *Applied meta-analysis with R*. CRC Press.

Cutting, J. C., & Ferreira, V. S. (1999). Semantic and phonological information flow in the production lexicon. *Journal of Experimental Psychology: Learning, Memory, and Cognition*, *25*(2), 318.

Damian, M. F., & Martin, R. C. (1999). Semantic and phonological codes interact in single word production. *Journal of Experimental Psychology. Learning, Memory, and Cognition, 25*(2), 345–361. https://doi.org/10.1037//0278-7393.25.2.345

Damian, M. F., & Bowers, J. S. (2003). Locus of semantic interference in picture-word interference tasks. *Psychonomic Bulletin & Review*, *10*(1), 111–117. https://doi.org/10.3758/BF03196474

Damian, M. F., & Spalek, K. (2014). Processing different kinds of semantic relations in picture-word interference with non-masked and masked distractors. *Frontiers in Psychology*, *5*. https://doi.org/10.3389/fpsyg.2014.01183

De Jong, R., Berendsen, E., & Cools, R. (1999). Goal neglect and inhibitory limitations: Dissociable causes of interference effects in conflict situations. *Acta psychologica*, *101*(2-3), 379-394.

De Jong, R., Liang, C. C., & Lauber, E. (1994). Conditional and unconditional automaticity: a dual-process model of effects of spatial stimulus-response correspondence. *Journal of Experimental Psychology: Human Perception and Performance*, *20*(4), 731.

de Zubicaray, G. I., Hansen, S., & McMahon, K. L. (2013). Differential processing of thematic and categorical conceptual relations in spoken word production. *Journal of*

*Experimental Psychology. General*, *142*(1), 131–142.
https://doi.org/10.1037/a0028717

Eriksen, B. A., & Eriksen, C. W. (1974). Effects of noise letters upon the identification of a target letter in a nonsearch task. *Perception & psychophysics*, *16*(1), 143-149.

Fisher, R. A. (1915). Frequency distribution of the values of the correlation coefficient in samples from an indefinitely large population. *Biometrika*, *10*(4), 507-521.

Finocchiaro, C., & Navarrete, E. (2013). About the locus of the distractor frequency effect: Evidence from the production of clitic pronouns. *Journal of Cognitive Psychology*, *25*(7), 861–872. https://doi.org/10.1080/20445911.2013.832254

Fuhrmeister, P., Madec, S., Lorenz, A., Elbuy, S., & Bürki, A. (submitted). Behavioural and EEG evidence for inter-individual variability in late encoding stages of word production.

Gauvin, H. S., Jonen, M. K., Choi, J., McMahon, K., & Zubicaray, G. I. de. (2018). No lexical competition without priming: Evidence from the picture–word interference paradigm: *Quarterly Journal of Experimental Psychology*. https://doi.org/10.1177/1747021817747266

Glaser, W. R., & Düngelhoff, F.-J. (1984). The time course of picture-word interference. *Journal of Experimental Psychology: Human Perception and Performance, 10*(5), 640–654. https://doi.org/10.1037/0096-1523.10.5.640

Hartendorp, M. O., Van der Stigchel, S., & Postma, A. (2013). To what extent do we process the nondominant object in a morphed figure? Evidence from a picture–word interference task. *Journal of Cognitive Psychology*, *25*(7), 843-860.

Hutson, J., & Damian, M. F. (2014). Semantic gradients in picture-word interference tasks: Is the size of interference effects affected by the degree of semantic overlap? *Frontiers in Psychology*, *5*. https://doi.org/10.3389/fpsyg.2014.00872

Janssen, N., Schirm, W., Mahon, B. Z., & Caramazza, A. (2008). Semantic interference in a delayed naming task: Evidence for the response exclusion hypothesis. *Journal of Experimental Psychology. Learning, Memory, and Cognition*, *34*(1), 249–256. https://doi.org/10.1037/0278-7393.34.1.249

Jongman, S. R., Roelofs, A., & Meyer, A. S. (2015). Sustained attention in language production: An individual differences investigation. *Quarterly Journal of Experimental Psychology*, *68*(4), 710-730.

Laganaro, M., Valente, A., & Perret, C. (2012). Time course of word production in fast and slow speakers: a high density ERP topographic study. *NeuroImage*, *59*(4), 3881-3888.

Lupker, S. J. (1979). The semantic nature of response competition in the picture-word interference task. *Memory & Cognition*, *7*(6), 485–495. https://doi.org/10.3758/BF03198265

Mädebach, A., Oppermann, F., Hantsch, A., Curda, C., & Jescheniak, J. D. (2011). Is there semantic interference in delayed naming? *Journal of Experimental Psychology. Learning, Memory, and Cognition*, *37*(2), 522–538. https://doi.org/10.1037/a0021970

Massidda, D. (2013). retimes: Reaction Time Analysis. R package version 0.1-2. https://CRAN.R-project.org/package=retimes

Piai, V., Roelofs, A., Jensen, O., Schoffelen, J.-M., & Bonnefond, M. (2014). Distinct Patterns of Brain Activity Characterise Lexical Activation and Competition in Spoken Word Production. *PLOS ONE*, *9*(2), e88674. https://doi.org/10.1371/journal.pone.0088674

Piai, V., Roelofs, A., & Schriefers, H. (2012). Distractor strength and selective attention in picture-naming performance. *Memory & Cognition*, *40*(4), 614–627. https://doi.org/10.3758/s13421-011-0171-3

Pratte, M. S., Rouder, J. N., Morey, R. D., & Feng, C. (2010). Exploring the differences in distributional properties between Stroop and Simon effects using delta plots. *Attention, Perception, & Psychophysics, 72*(7), 2013-2025. doi:10.3758/APP.72.7.2013

Proctor, R. W., Miles, J. D., & Baroni, G. (2011). Reaction time distribution analysis of spatial correspondence effects. *Psychonomic Bulletin & Review, 18*(2), 242-266.

R Core Team (2020). R: A language and environment for statistical computing. R Foundation for Statistical Computing, Vienna, Austria. URL https://www.R-project.org/.

Ridderinkhof, K. R., Scheres, A., Oosterlaan, J., & Sergeant, J. A. (2005). Delta plots in the study of individual differences: new tools reveal response inhibition deficits in AD/Hd that are eliminated by methylphenidate treatment. *Journal of abnormal psychology*, *114*(2), 197.

Ridderinkhof, K. R., van den Wildenberg, W. P., Wijnen, J., & Burle, B. (2004). Response inhibition in conflict tasks is revealed in delta plots. *Cognitive neuroscience of attention*, *369*, 377.

Rodríguez-Ferreiro, J., Davies, R., & Cuetos, F. (2014). Semantic domain and grammatical class effects in the picture–word interference paradigm. *Language, Cognition and Neuroscience*, *29*(1), 125–135. https://doi.org/10.1080/01690965.2013.788195

Roelofs, A. (2008). Dynamics of the attentional control of word retrieval: Analyses of response time distributions. *Journal of Experimental Psychology: General*, *137*(2), 303.

Roelofs, A., & Piai, V. (2017). Distributional analysis of semantic interference in picture naming. *Quarterly Journal of Experimental Psychology*, *70*(4), 782-792.

Roelofs, A., Piai, V., & Garrido Rodriguez, G. (2011). Attentional inhibition in bilingual naming performance: evidence from delta-plot analyses. *Frontiers in psychology*, *2*, 184.

Sailor, K., Brooks, P. J., Bruening, P. R., Seiger-Gardner, L., & Guterman, M. (2009). Exploring the time course of semantic interference and associative priming in the picture–word interference task: *Quarterly Journal of Experimental Psychology*. https://doi.org/10.1080/17470210802254383

Scaltritti, M., Navarrete, E., & Peressotti, F. (2015). Distributional analyses in the picture–word interference paradigm: Exploring the semantic interference and the distractor frequency effects. *The Quarterly Journal of Experimental Psychology*, *68*(7), 1348–1369. https://doi.org/10.1080/17470218.2014.981196

Schad, D. J., Vasishth, S., Hohenstein, S., & Kliegl, R. (2020). How to capitalize on a priori contrasts in linear (mixed) models: A tutorial. *Journal of Memory and Language*, *110*, 104038.

Shao, Z., Roelofs, A., & Meyer, A. S. (2012). Sources of individual differences in the speed of naming objects and actions: The contribution of executive control. *Quarterly Journal of Experimental Psychology*, *65*(10), 1927-1944.

Shao, Z., Meyer, A. S., & Roelofs, A. (2013). Selective and nonselective inhibition of competitors in picture naming. *Memory & cognition*, *41*(8), 1200-1211.

Shao, Z., Roelofs, A., Acheson, D. J., & Meyer, A. S. (2014). Electrophysiological evidence that inhibition supports lexical selection in picture naming. Brain Research, 1586, 130-142.

Shao, Z., Roelofs, A., Martin, R. C., & Meyer, A. S. (2015). Selective inhibition and naming performance in semantic blocking, picture-word interference, and color–word Stroop tasks. *Journal of Experimental Psychology: Learning, Memory, and Cognition*, *41*(6), 1806.

Signorell, A. et mult. al. (2020). DescTools: Tools for descriptive statistics. R package version 0.99.38.

Sutton, A. J., & Abrams, K. R. (2001). Bayesian methods in meta-analysis and evidence synthesis. *Statistical methods in medical research*, *10*(4), 277-303.

Starreveld, P. A., & La Heij, W. (1996). Time-course analysis of semantic and orthographic context effects in picture naming. *Journal of Experimental Psychology: Learning, Memory, and Cognition, 22*(4), 896–918. https://doi.org/10.1037/0278-7393.22.4.896

Van Den Wildenberg, W. P., Wylie, S. A., Forstmann, B. U., Burle, B., Hasbroucq, T., & Ridderinkhof, K. R. (2010). To head or to heed? Beyond the surface of selective action inhibition: a review. *Frontiers in human neuroscience*, *4*, 222.

Van Maanen, L., & Van Rijn, H. (2008). The picture-word interference effect is a Stroop effect after all. In *Proceedings of the Annual Meeting of the Cognitive Science Society* (Vol. 30, No. 30).

Vieth, H. E., McMahon, K. L., & de Zubicaray, G. I. (2014). The roles of shared vs. Distinctive conceptual features in lexical access. *Frontiers in Psychology*, *5*. https://doi.org/10.3389/fpsyg.2014.01014

Wetzels, R., & Wagenmakers, E. J. (2012). A default Bayesian hypothesis test for correlations and partial correlations. Psychonomic bulletin & review, 19(6), 1057-1064. https://doi.org/10.3758/s13423-012-0295-x

Zhang, Q., Feng, C., Zhu, X., & Wang, C. (2016). Transforming semantic interference into facilitation in a picture–word interference task. *Applied Psycholinguistics*, *37*(5), 1025-1049.

Zhang, J., & Kornblum, S. (1997). Distributional analysis and De Jong, Liang, and Lauber's (1994) dual-process model of the Simon effect. *Journal of Experimental Psychology: Human Perception and Performance, 23*(5), 1543. https://doi.org/10.1037/0096-1523.23.5.1543

Appendix A. Sensitivity analyses.

1. Correlation between the magnitude of the semantic interference effect and the slowest delta plot segment (participant analysis)

| Prior | Estimate | 95% CrI | tau | 95% CrI | BF₁₀ |
|---|---|---|---|---|---|
| **Uniform (-3,3)** | .52 | [.44, .60] | .23 | [.16, .31] | 451430705865945 |
| **Uniform (-10,10)** | .52 | [.43, .60] | .23 | [.16, .31] | 139367947015877 |

2. Correlation between the magnitude of the semantic interference effect and the fastest delta plot segment (participant analysis)

| Prior | Estimate | 95% CrI | tau | 95% CrI | BF₁₀ |
|---|---|---|---|---|---|
| **Uniform (-3,3)** | .18 | [.11, .24] | .14 | [.04, .22] | 544 |
| **Uniform (-10,10)** | .18 | [.10, .25] | .14 | [.03, .22] | 163 |

3. Magnitude of the semantic interference effect in first quintile

| Prior | Estimate | 95% CrI | tau | 95% CrI | BF₁₀ |
|---|---|---|---|---|---|
| **Normal (0,200)** | 6 ms | [3, 9] | 2 ms | [0, 5] | 294 |
| **Uniform (-200,200)** | 6 ms | [3, 9] | 2 ms | [0, 5] | 374 |
| **Normal (0,50)** | 6 ms | [3, 9] | 2 ms | [0, 5] | 1184 |

4. Magnitude of the semantic interference effect in second quintile

| Prior | Estimate | 95% CrI | tau | 95% CrI | BF₁₀ |
|---|---|---|---|---|---|
| **Normal (0,200)** | 11 ms | [8, 14] | 6 ms | [3, 9] | 11302332 |
| **Uniform (-200,200)** | 11 ms | [8, 14] | 6 ms | [3, 9] | 14409809 |
| **Normal (0,50)** | 11 ms | [8, 14] | 6 ms | [3, 9] | 44616986 |

5. Magnitude of the semantic interference effect in third quintile

| Prior | Estimate | 95% CrI | tau | 95% CrI | BF₁₀ |
|---|---|---|---|---|---|
| **Normal (0,200)** | 17 ms | [13, 21] | 10 ms | [7,14] | 1014809713 |
| **Uniform (-200,200)** | 17 ms | [13, 21] | 10 ms | [7,14] | 1305691783 |
| **Normal (0,50)** | 17 ms | [13, 21] | 10 ms | [7,14] | 3958482766 |

6. Magnitude of the semantic interference effect in fourth quintile

| Prior | Estimate | 95% CrI | tau | 95% CrI | BF$_{10}$ |
|---|---|---|---|---|---|
| **Normal (0,200)** | 27 ms | [22, 32] | 15 ms | [11, 20] | 189253776744 |
| **Uniform (-200,200)** | 27 ms | [22, 32] | 15 ms | [11, 20] | 232555810304 |
| **Normal (0,50)** | 27 ms | [21, 32] | 15 ms | [11, 20] | 664096312968 |

7. Magnitude of the semantic interference effect in fifth quintile

| Prior | Estimate | 95% CrI | tau | 95% CrI | BF$_{10}$ |
|---|---|---|---|---|---|
| **Normal (0,200)** | 49 ms | [39, 58] | 21 ms | [12, 32] | 125957326795 |
| **Uniform (-200,200)** | 49 ms | [39, 58] | 22 ms | [12, 32] | 163257419459 |
| **Normal (0,50)** | 48 ms | [39, 58] | 21 ms | [12, 32] | 338387585153 |

8. Correlation between the magnitude of the semantic interference effect and the slowest delta plot segment (item analysis)

| Prior | Estimate | 95% CrI | tau | 95% CrI | BF$_{10}$ |
|---|---|---|---|---|---|
| **Uniform (-3,3)** | .49 | [.43, .54] | .07 | [.01, .15] | 21345721438660358144 |
| **Uniform (-10,10)** | .49 | [.43, .54] | .07 | [.00, .15] | 6310900996115787776 |

9. Correlation between the magnitude of the semantic interference effect and the fastest delta plot segment (item analysis)

| Prior | Estimate | 95% CrI | tau | 95% CrI | BF$_{10}$ |
|---|---|---|---|---|---|
| **Uniform (-3,3)** | .31 | [.26, .37] | .04 | [.00, .12] | 12428222347230 |
| **Uniform (-10,10)** | .31 | [.26, .37] | .04 | [.00, .12] | 3633637707907 |

Appendix B. Studies used in meta-analyses.

| Experiments | Reference |
| --- | --- |
| **Aristei.2011** | Aristei et al. (2011) |
| **Aristei.2013** | Aristei & Abdel Rahman (2013) |
| **Cutting.1999.1** | Cutting & Ferreira (1999) |
| **Cutting.1999.2** | |
| **Cutting.1999.3a.1** | |
| **Cutting.1999.3a.2** | |
| **Damian.2003.SOA-100** | Damian & Bowers (2003) |
| **Damian.2003.SOA0** | |
| **Damian.2003.SOA100** | |
| **Damian.2014** | Damian & Spalek (2014) |
| **deZubicaray.2013** | De Zubicaray et al. (2013) |
| **Finocchiaro.2013.1** | Finocchiaro & Navarrete (2013) |
| **Fuhrmeister.unpublished** | - |
| **Fuhrmeister2.unpublished** | - |
| **Gauvin.2018.1.fam** | Gauvin et al. (2018) |
| **Gauvin.2018.1.nofam** | |
| **Gauvin.2018.2.fam** | |
| **Gauvin.2018.2.nofam** | |
| **Hartendorp.2013.1** | Hartendorp et al. (2013) |
| **Hartendorp.2013.2** | |
| **Hutson.2014.1** | Hutson & Damian (2014) |
| **Hutson.2014.2** | |
| **Janssen.2008.1a** | Janssen et al. (2008) |
| **Janssen.2008.2a** | |
| **Maedebach.2011.2** | Mädebach et al. (2011) |
| **Maedebach.2011.4** | |
| **Maedebach.2011.5a** | |
| **Maedebach.2011.6** | |
| **Piai.2012** | Piai et al. (2012) |
| **Piai.2012.2** | |
| **Piai.2014** | Piai et al. (2014) |
| **Piai.unpublished** | - |
| **Python.unpublished** | - |
| **Rodriguez.2014** | Rodríguez-Ferreiro et al. (2014) |
| **Roelofs.2008.3** | Roelofs (2008) |
| **Sailor.2009.1.SOA-150** | Sailor et al. (2009) |
| **Sailor.2009.1.SOA150** | |
| **Sailor.2009.2.SOA0** | |
| **Scaltritti.2015.1** | Scaltritti et al. (2015) |
| **Scaltritti.2015.3** | |
| **Shao.2015.Exp.1** | Shao et al. (2015) |
| **Shao.2015.Exp.2** | |
| **Starreveld.2013.1.SOA0** | Starreveld et al. (2013) |
| **Starreveld.2013.1.SOA43** | |
| **Starreveld.2013.1.SOA86** | |
| **Starreveld.2013.1.SOAminus43** | |
| **Starreveld.2013.1.SOAminus86** | |
| **vanRijn.unpublished** | - |
| **Vieth.2014.SOA-160** | Vieth et al. (2014) |
| **Vieth.2014.SOA0** | |
| **Zhang.2016.1.SOA-100** | Zhang et al. (2016) |
| **Zhang.2016.1.SOA0** | |
| **Zhang.2016.1.SOA100** | |
| **Zhang.2016.2.SOA0** | |